# A model of electrical impedance tomography on peripheral nerves for a neural-prosthetic control interface

J. Hope[1,2], F. Vanholsbeeck[2], A. McDaid[1]

[1]Department of Mechanical Engineering, The University of Auckland, 5 Grafton Road, Auckland 1010, NZ
[2]Dodd Walls Centre, The Department of Physics, The University of Auckland, 38 Princes Street, Auckland 1010, NZ

**Abstract**

**Objective:** A model is presented to evaluate the viability of using electrical impedance tomography (EIT) with a nerve cuff to record neural activity in peripheral nerves.
**Approach:** Established modelling approaches in neural-EIT are expanded on to be used, for the first time, on myelinated fibres which are abundant in mammalian peripheral nerves and transmit motor commands.
**Main results:** Fibre impedance models indicate activity in unmyelinated fibres can be screened out using operating frequencies above 100 Hz. At 1 kHz and 10 mm electrode spacing, impedance magnitude of inactive intra-fascicle tissue and the fraction changes during neural activity are estimated to be 1,142 $\Omega$.cm and -8.8x10$^{-4}$, respectively, with a transverse current, and 328 $\Omega$.cm & -0.30, respectively with a longitudinal current. We show that a novel EIT drive and measurement electrode pattern which utilises longitudinal current and longitudinal differential boundary voltage measurements could distinguish activity in different fascicles of a three-fascicle mammalian nerve using pseudo-experimental data synthesised to replicate real operating conditions.
**Significance:** The results of this study provide an estimate of the transient change in impedance of intra-fascicle tissue during neural activity in mammalian nerve, and present a viable EIT electrode pattern, both of which are critical steps towards implementing EIT in a nerve cuff for neural prosthetics interfaces.

## I. INTRODUCTION

Limb loss is projected to affect 2.2 million people in the USA by 2020, of which 35% of cases affect the upper limb (Ziegler-Graham, MacKenzie et al. 2008). Prosthetics offer a means to improve quality of life for amputees by restoring the lost functions needed to perform activities of daily life (Spiers, Resnik et al. 2017). However, modern electric prosthetics for transradial and transhumeral amputees contain more mechanical degrees of freedom than can be intuitively controlled due to limitations in the neural-interface (Graimann and Dietl 2013). Increased dexterity/ functionality and increased control are two factors which are desired by users (Biddiss, Beaton et al. 2007, Engdahl, Chestek et al. 2017) and contribute to lowering user rejection rates (Østlie, Lesjø et al. 2012). A study by (Engdahl, Chestek et al. 2017) reported 64% of users are interested to test an implantable peripheral nerve interface (PNI) if it improves prosthetic functionality despite the medical risks associated with some PNI hardware.

Among neural interface technologies a trade-off exists between beneficial characteristics, with no single device achieving high chronic stability, a high number of control channels (selectivity) and low invasiveness (Ortiz-Catalan, Brånemark et al. 2012, del Valle and Navarro 2013, Thompson, Zoratti et al. 2016). Nerve cuffs are the least invasive and least selective peripheral nerve interface. Researchers have used nerve cuffs to identify touch force and slip (Haugland and Hoffer 1994, Haugland, Hoffer et al. 1994), estimate joint angle (Chan, Lin et al. 2012), and as a sensor in closed loop functional electrical stimulation (FES) systems (Haugland and Sinkjaer 1995, Haugland, Lickel et al. 1999, Inmann and Haugland 2004). While many researchers have shown it is possible to classify multiple concurrent signals based on the type of sensory information (Raspopovic, Carpaneto et al. 2010, Rieger and Taylor 2013, Al-Shueli, Clarke et al. 2014, Karimi and Seydnejad 2015), it has not been possible to differentiate more than two concurrent signals of the same type (Raspopovic, Carpaneto et al. 2010, Zariffa 2015), which is a major barrier to their use in prosthetic limbs.

Electrical Impedance Tomography (EIT) is an imaging modality with electrical impedance as the contrast agent and biomedical applications in monitoring of respiratory and pulmonary systems (Frerichs, Amato et al.

2017), identification of ischaemic brain tissue for diagnosis of stroke (Dowrick, Blochet et al. 2015), and localisation of epileptic foci (Dowrick, Dos Santos et al. 2015, Aristovich, Packham et al. 2016) amongst others (Bayford, Ollmar et al. 2017). The application of EIT to a nerve cuff offers a potential means to classify multiple concurrent motor-command signals based on their location within the peripheral nerve. Somatotopic organisation of fibres and fascicles, which exists to some degree in peripheral nerves (Jabaley 1980, Tyler, Polasek et al. 2015), gives promise to a tomographic approach to nerve cuffs for control of prosthetic limbs. In differential EIT, transient changes in the impedance distribution within the sample acts as the contrast agent. The change in impedance of neural tissue during activity in neural cells is attributed to the increased membrane conductivity during action potentials (Holder 1992, Boone 1995). Neural-EIT has been developed using nerve fibre impedance models based on unmyelinated fibre membrane dynamics and single cable theory, and a crab nerve animal model in experiments (Holder 1992, Boone 1995, Liston 2003, Gilad, Ghosh et al. 2009, Oh, Gilad et al. 2011, Liston, Bayford et al. 2012, Aristovich, Dos Santos et al. 2015). Nerve fibre impedance models have been combined with physiological information on the mammalian brain to image cortical activity during sensory stimuli in humans (Davidson, Wright et al. 2010, Pomfrett, Davidson et al. 2010, McCann, Ahsan et al. 2011, Pollard, Pomfrett et al. 2011) and in rats (Aristovich, Santos et al. 2014, Vongerichten 2015, Aristovich, Packham et al. 2016, Vongerichten, dos Santos et al. 2016). In rats, spatial and temporal resolutions of 0.05mm & 2ms (Aristovich, Santos et al. 2014, Aristovich, Packham et al. 2016) and 0.4mm & 3.3ms (Vongerichten 2015, Vongerichten, dos Santos et al. 2016) were achieved, which if replicated on a nerve cuff would provide a significant improvement in nerve cuff selectivity. Efforts to apply EIT to a nerve cuff have reported transient longitudinal impedance changes in the nerve (Fouchard, Coizet et al. 2017) , fascicle level selectivity in rat sciatic nerve (Aristovich 2016), and spontaneous activity of cardiac and pulmonary functions in vagus nerve of sheep (Aristovich 2017).

In this study, we investigate the application of EIT to a nerve cuff for control of a robotic prosthetic. We applied, for the first time, modelling approaches for neural-EIT to myelinated fibres. This modelling produced estimates of anisotropic impedance of intra-fascicle tissue (fibres and extracellular space) which allows us to evaluate potential EIT drive and measurement electrode patterns. We investigate the expected performance of a novel EIT electrode pattern, implemented on a 32 channel dual-ring electrode array, for EIT reconstruction of neural activity in a multi-fascicle mammalian nerve using pseudo-experimental data synthesised to replicate real operating conditions. The results provide insights into operating parameters and potential EIT electrode patterns in an experimental set-up, and the viability of a nerve-cuff based EIT interface for neural prosthetics.

## II. METHODS

In accordance with other research we separated our model into three components (Boone and Holder 1995, Oh, Gilad et al. 2011, Liston, Bayford et al. 2012, Aristovich, Santos et al. 2014). The first component calculates transient changes to the conductivity of the nerve fibre membrane during propagation of an action potential using a Hodgkin-Huxley type model of ion channel behaviour (Hodgkin and Huxley 1952) or using empirical data. We extended existing models to account for myelinated fibres by using the Richardson model of the node of Ranvier (Richardson, McIntyre et al. 2000). The second component calculates the specific impedance of nerve fibre under an oscillating extracellular current by modelling the smallest repeating unit of a fibre in extracellular space with an equivalent electrical circuit, and then solving this model with cable-theory. We modified this for myelinated fibres by evaluating a single-cable model with lumped parameters (Richardson, McIntyre et al. 2000), and a more physiologically accurate double cable model with non-uniform ion channel distribution (Halter and Clark 1991, Scherer and Arroyo 2002). We have split this second component into two sections in this paper, in order to address the transverse and longitudinal impedances separately. The third component uses bulk tissue conductivities in a layered volume conductor model of the nerve which is solved using the finite element method (FEM) to estimate boundary voltage data from given tissue conductivity distributions. We based the morphology of our model on a three-fascicle rat sciatic nerve (Layton and Sastry 2004, Prodanov 2006).

The rat sciatic nerve offers similarities in morphology to the human median nerve. The rat sciatic nerve has 3 to 4 fascicles ranging in size from 0.05 to 1mm$^2$ (from our own histology (Hope 2017)) each containing a heterogeneous distribution of motor fibres ranging in size from 10 to 15µm (Prodanov 2006). In comparison, the median nerve in the human upper limb contains 10 or more fascicles at the elbow, which range in size from 0.12 to 2mm$^2$ each containing a heterogeneous distribution of motor fibres ranging in size from 12 to 22µm (Delgado-

Martínez, Badia et al. 2016). All three major nerves which traverse the elbow in the human upper limb, the median, ulnar and radial, contain multi-fascicle morphology (Sunderland 1945) and innervate muscles which control the motor functions of the hand and wrist (Jabaley, Wallace et al. 1980).

*A. Membrane Conductivity*

For unmyelinated fibres, we calculated the membrane conductivity using the parameters and equations for rats and mice provided in a 2014 study (Neishabouri and Faisal 2014). In myelinated fibres, the ion channels on the node of Ranvier control the majority of the transmembrane ion-flux during the depolarisation phase of an action potential. Membrane conductivity at the node of Ranvier was calculated using parameters and equations from (Brazhe, Maksimov et al. 2011) as the sum of the fast Sodium, persistent Sodium, and slow Potassium ion channels, as well as the passive membrane leakage.

*B. Transverse Fibre Impedance*

Impedance models of an unmyelinated fibre are derived from assumptions that the membrane is a semi-conductive thin-walled cylinder of constant diameter which separates two conductive volumes, the intracellular fluid and extracellular space. In transverse orientation, the electric current is applied perpendicular to the fibre length, and, accordingly, in longitudinal orientation, the current is applied parallel to the fibre length.

The specific impedance of an unmyelinated fibre in a transverse current $z_{U,t}$ (Ωm) was calculated using equation 1 (Bozler and Cole 1935, Boone 1995):

$$z_{U,t} = r_e \frac{r_e(1-\alpha)+(r_i + z_m/a_m)(1+\alpha)}{r_e(1+\alpha)+(r_i + z_m/a_m)(1-\alpha)} \qquad (1)$$

where $\alpha$ is the volume fraction of the fibre in extracellular space, $a_m$ the membrane radius, $r_e$ extracellular space resistivity (Ωm), and $r_i$ intracellular fluid resistivity (Ωm). Membrane impedance is given by $z_m = \frac{r_m}{1+j\omega r_m c_m}$, where j is the imaginary operator, ω the angular frequency (rad/s), $r_m$ membrane area resistivity (Ωm$^2$), $c_m$ membrane area capacitance (Fm$^{-2}$).

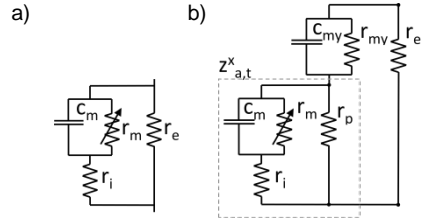

Figure 1: Transverse current equivalent electrical circuits of unmyelinated fibres and the node of Ranvier region in myelinated fibres (a), and the paranode, juxtaparanode and internode regions of myelinated fibres (b). In both models an oscillating current is applied across the extracellular resistance $r_e$ to determine the circuit impedance.

The presence of Schwann cells around a myelinated fibre axon and variable ion channel concentration creates a more complicated cytostructure which must be reflected in the equivalent electrical circuits. The smallest repeating unit of a myelinated fibre extends between two neighbouring nodes of Ranvier, and can be broken down into four regions as described in (McIntyre, Richardson et al. 2002, Scherer and Arroyo 2002); these are 1) the node of Ranvier, which primarily contains fast Sodium ion channels; 2) the paranode, where the myelin sheath attaches to the axon; 3) the juxtaparanode which primarily contains fast Potassium and slow Potassium ion channels; and 4) the internode, with a relatively low concentration of ion channels.

The specific impedance of a myelinated fibre in the presence of a transverse current $z_{M,t}$ was calculated by summing the impedances of each region together in a 'parallel' configuration after weighting the values by their length fraction. Each length fraction was calculated as the region length divided by the node-to-node distance, using dimensional data from (McIntyre, Richardson et al. 2002). Specific transverse impedance of the node of

Ranvier $z_{M,t}^n$ was calculated using equation 1. For the remaining regions, the specific impedance was calculated by first modifying the parameters in equation 1 to describe the axon in Periaxonal space:

$$z_{a,t}^x = r_p \frac{r_p(1-\alpha_a)+\left(r_i+ {Z_m}/{a_m}\right)(1+\alpha_a)}{r_p(1+\alpha_a)+\left(r_i+ {Z_m}/{a_m}\right)(1-\alpha_a)} \qquad (2)$$

where superscript $x$ indicates the region, $r_p$ is the periaxonal space resistivity, and $\alpha_a$ is the volume fraction of axon in periaxonal space. The axon impedance was then nested inside a myelin sheath and extracellular space, see Fig 1, by again modifying the parameters in equation 1:

$$z_{M,t}^x = r_e \frac{r_e(1-\alpha_{my})+\left(z_{a,t}^x+ {z_{my}}/{\tilde{a}_{my}}\right)(1+\alpha_{my})}{r_e(1+\alpha_{my})+\left(z_{a,t}^x+ {z_{my}}/{\tilde{a}_{my}}\right)(1-\alpha_{my})} \qquad (3)$$

where $\tilde{a}_{my}$ is the average myelin radius, $z_{my} = \frac{r_{my}}{1+j\omega r_{my} c_{my}}$ as before but with the subscript my denoting myelin sheath properties, and $\alpha_{my}$ is the volume fraction of the axon and periaxonal space inside the myelin and extracellular space. Current is assumed to only flow radially within the myelin sheath and so the volume occupied by the myelin sheath was omitted when calculating $\alpha_{my}$.

*C. Longitudinal Fibre Impedance*

The specific impedance of an unmyelinated fibre in a longitudinal current $z_{U,l}$ was calculated using equation 4 (Liston, Bayford et al. 2012):

$$z_{U,l} = \frac{r_i r_e}{[(1-\alpha)r_e+\alpha r_i]} + \frac{r_e^2(1-\alpha)}{\alpha[(1-\alpha)r_e+\alpha r_i](LF)} e^{-LF} \sinh(LF) \qquad (4)$$

where the electrodes are separated along the fibre length by distance $2L$, and the parameter $F = \frac{\sqrt{1+j\omega\tau}}{\lambda}$ contains the time constant $\tau = r_m c_m$ and length constant $\lambda = \sqrt{\frac{a r_m}{2\left(r_i+r_e\frac{(1-\alpha)}{\alpha}\right)}}$ from the standard cable equation, used to model the fibre.

The specific impedance of a myelinated fibre in the presence of a longitudinal current $z_{M,l}$ was calculated using two models. The first, termed the 'physiological model', segmented the fibre length into the four types of regions described earlier in section *B. Transverse Fibre Impedance*. Each region was separated into five layers, see Fig 2a Three of the five layers were longitudinal current paths: the extracellular space, periaxonal space, and intracellular space. The remaining two layers were transverse current paths through the myelin and the axon membrane, which connected the longitudinal current paths and created a more physiologically accurate double cable type model.

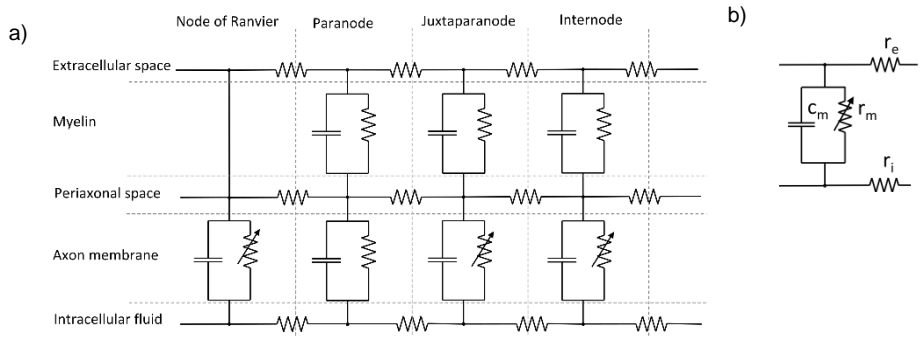

Figure 2: Part of the longitudinal current equivalent electrical circuit of the physiological model of a myelinated fibre (a) showing the five layers (extracellular space, myelin, periaxonal space, axon membrane, and intracellular fluid) and the four regions (node of Ranvier, paranode, juxtaparanode, and internode). Longitudinal current equivalent electrical circuit used for an unmyelinated fibre and for the lumped parameter model of the myelinated fibre (b). In both equivalent electrical circuits (a) & (b), an oscillating current is applied across the extracellular resistance to determine the circuit impedance.

The second longitudinal fibre model, termed the 'lumped parameter model', omitted the Periaxonal space and the axon membrane in the myelinated regions, creating a single cable model. In addition, the myelin layer in each region and the node of Ranvier membrane were represented, respectively, by a single resistor and a single capacitor in parallel, producing the same equivalent electrical circuit as was used in unmyelinated fibres, Fig 2b.

Both longitudinal current models were built in the electronic circuit analysis software *LTspice* for a length of fibre extending between two neighbouring nodes of Ranvier using absolute resistance and capacitance values calculated for a 5.7µm diameter myelinated fibre from electrical properties in table 1 and dimensions from (McIntyre, Richardson et al. 2002). The impedance of the single cable model was evaluated using equation 4.

*D. Nerve Tissues*

The change in the four fibre impedance parameters during neural activity ($\Delta z_{U,t}$, $\Delta z_{U,l}$, $\Delta z_{M,t}$, and $\Delta z_{M,l}$) were calculated using the two-state model approach outlined in (Liston, Bayford et al. 2012). The two state model calculates the fibre impedance in the active state using the maximum axon membrane conductivity attained during an action potential and then calculates the fraction change in impedance $\Delta z_0$ using $\Delta z_0 = \frac{z_{x,y}^d - z_{x,y}^p}{z_{x,y}^p}$, where subscripts $x$ and $y$ indicate the fibre type and orientation respectively, and superscripts $d$ and $p$ indicate the active (depolarised) and inactive (polarised) states respectively. The axon membrane conductivity in the active and inactive fibre states for unmyelinated fibres and the node of Ranvier region of myelinated fibres were calculated as described earlier in *section A*. In myelinated fibre the axon membrane of the paranode region was assumed to have constant conductivity (McIntyre, Richardson et al. 2002), and in the juxtaparanode and internode regions axon membrane conductivity was calculated using ion channel concentration values from (Brazhe, Maksimov et al. 2011).

The fraction change in impedance of tissue in the intra-fascicle volume $\Delta z_{IFV}$ was calculated by separating the range of fibre diameters present in the nerve into $n$ sub-groups and summing them as in a 'parallel' configuration:

$$\frac{1}{\Delta z_{IFV}} = \sum_{i=1}^{n} \frac{T_{p,i} \, C_{p,i} \, V_i}{\Delta z_{0,i}} \qquad (5)$$

where subscript $i$ indicates the $i^{th}$ fibre diameter sub-group, $T_p$ and $C_p$ are factors which correct for, respectively, the axon membrane capacitance and the fibre diameter dependent signal velocities. The volume fraction $V$ was calculated from fibre diameter population data for rat sciatic nerve in (Prodanov 2006).

The capacitance correction factor $T_p$ was calculated as the difference in the maximum change in voltage between two parallel RC circuits as described in Ref (Liston, Bayford et al. 2012). One circuit had zero capacitance and the other a capacitance value taken from the lumped parameter fibre-model. For both circuits, the resistance changed as a function of time between that of the inactive and the active fibre states, with values again taken from the lumped parameter model. In myelinated fibres, the time varying function used was the falling edge of a sinusoidal signal with 12.5 kHz frequency, which was intended to approximate the 40µs rise time in membrane conductivity of the node of Ranvier during the depolarisation phase of an action potential. A sinusoid of 250Hz frequency was used to approximate the repolarisation phase. In unmyelinated fibres, the frequencies used for the depolarisation and repolarisation phases were 250 and 100 Hz respectively to approximate a 2ms rise time and 5ms fall time of the membrane conductivity during an action potential.

The signal velocity correction factor $C_p$ was calculated as the product of signal dispersion effects and signal transit time effects. Signal dispersion assumed 30 mm between the source of the compound action potential and the nerve cuff. Signal transit times through the nerve cuff accounted for the transient conductivity across the nodes of Ranvier within the nerve cuff. For longitudinal current an EIT imaging volume 10mm in length was assumed; in transverse current 2 mm length imaging volume was assumed.

Impedances of the passive tissues perineurium, epineurium and endoneurium obtained from mammalian peripheral nerves have not been reported before. Instead, it is common (as in (Choi, Cavanaugh et al. 2001, Yoo and Durand 2005, Kuhn 2008, Taghipour-Farshi, Frounchi et al. 2015, Garai, Koh et al. 2017)) to use value for perineurium (47,600 $\Omega$.cm) obtained from frog (Weerasuriya, Spangler et al. 1984), and for epineurium (1,211 $\Omega$.cm) from the transverse impedance of the dorsal column of cat at 1 kHz (Ranck Jr and BeMent 1965).

Endoneurium occupies the majority of the extracellular space within the intra-fascicle volume, with the remainder occupied by an approximately 40 – 100 nm thick layer of basal lamina immediately surrounding each fibre. The thickness of basal lamina was calculated from analysis of images in (Morris, Hudson et al. 1972, Geada

Trigo Calheiros De Figueiredo 2014, Pannese 2015). Values in literature for the resistivity of extracellular space vary between 175 and 1,000 Ω.cm (Frijns and ten Kate 1994, Richardson, McIntyre et al. 2000, Kuhn 2008, Zariffa 2009). We adopted an isotropic resistivity of 1,000 Ω.cm which, using equations 1 – 4 at 1 kHz and 10mm electrode spacing, produced transverse and longitudinal resistivity values which were close to those reported in Ref (Ranck Jr and BeMent 1965) from the dorsal column of cat of 1,211 Ω.cm radial and 138 to 217 Ω.cm longitudinal.

*E. Electrical Impedance Tomography*

The forward EIT problem is defined as $\mathbf{g}_n = \mathbf{A}\mathbf{c}_n$ where $\mathbf{A}$ is the sensitivity matrix, and in neural EIT $\mathbf{g}_n$ is the normalised change in boundary voltages and $\mathbf{c}_n$ the normalised conductivity change (Holder 2004). The sensitivity matrix is populated from boundary voltages generated from known currents in a Finite Element Method (FEM) model simulation of the system (Holder 2004, Aristovich, Santos et al. 2014). The FEM model requires knowledge of the size, shape and location of the electrode array, and, if possible, some a-priori information on the conductivity distribution of the sample. Our nerve cuff electrode array contained a total of 32 electrodes, each 1.1 x 0.11mm, arranged into two rings of 16 electrodes on a 22.5º pitch around the circumference, Fig 3. For our sample volume we used a single fascicle model of a 50mm length by 600 µm diameter peripheral nerve comprised of four cylindrical layers: an 80 µm thick saline fluid layer between the electrodes and nerve outer boundary, 50 µm thick epineurium layer (Dumanian, McClinton et al. 1999), 5 µm perineurium layer (Layton and Sastry 2004), and a 545 µm radius intra-fascicle volume in the centre. We divided the intra-fascicle volume into a grid of 49 sub-volumes which extended the length of the nerve, Fig 3. A single fascicle model is akin to assuming no a-priori information about the intra-fascicle structure of the nerve.

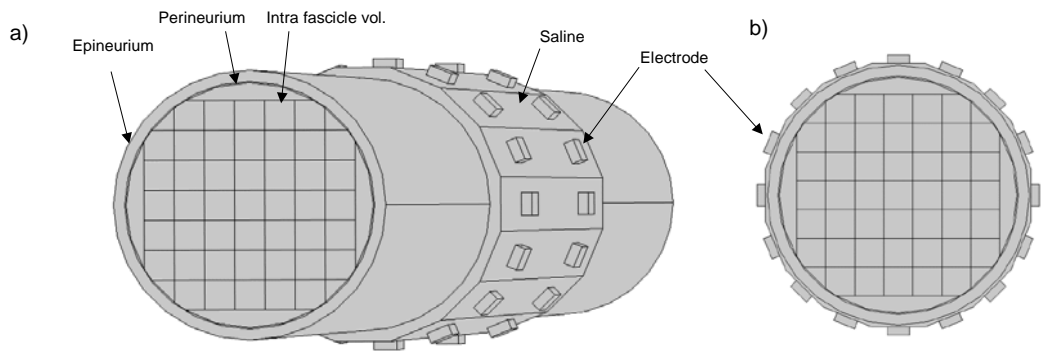

Figure 3: Two rings, each containing 16 electrodes, are spaced apart along the nerve length and wrapped around the cylindrical shell model of the nerve with layers for the saline fluid, epineurium, perineurium, and intra-fascicle sub-volumes visible in an angled end-and-side view of the FEM model (a) and in an end view of the FEM model (b) which has all 16 electrodes of one electrode ring visible.

We imposed a limit of one drive current pattern within the nerve cuff due to the millisecond to sub-millisecond time resolution required to resolve compound action potentials (Aristovich, Packham et al. 2016, Vongerichten, dos Santos et al. 2016) and the necessity to repeat the drive current pattern for all electrode pairs in each time resolution unit.

The nerve cuff with dual-ring electrode array can pass current between two drive electrodes on the same electrode-ring to generate transverse current patterns, or pass current between two drive electrodes on different electrode-rings to generate longitudinal current patterns, Fig 4. Simulations by (Graham and Adler 2007) compared seven possible drive current patterns within a dual-ring electrode array around a cylindrical sample volume with isotropic conductivity features for 3D EIT lung imaging of the thorax, concluding that transverse current pattern with drive current and voltage measurement between adjacent electrodes provided the best performance. Transverse current in a nerve would largely eliminate the anisotropy by operating in a plane perpendicular to the axis containing the unique conductivity, which we term the 'anisotropic axis', an approach adopted by (Aristovich 2016) on peripheral nerve and (Silva, Sousa et al. 2014) on muscle tissue.

Longitudinal current, on the other hand, in the dual-ring nerve cuff requires a 3D model and consideration of tissue anisotropy. Anisotropic conductivity anomalies produce boundary voltage data with non-unique solution

(Adler, Gaburro et al. 2015), however, numerical methods with some a-priori information have proven capable of reconstructing anisotropic anomalies in 2D simulations (Hamilton, Lassas et al. 2014, Adler, Gaburro et al. 2015, Wang, Xu et al. 2015) and in 3D simulations to manage anisotropy of white matter in the brain (Turovets, Volkov et al. 2014).

The unique combination of factors in our sample – differential 3D-EIT using dual-ring electrode array to detect conductivity variation in an anisotropic axis perpendicular to the planes of the electrode-rings – means there is no precedent for EIT drive and measurement electrode pattern. We propose a novel approach: to assume that transient conductivity variations of all tissues are uniform along the nerve's length dimension, which is the anisotropic axis in intra-fascicle tissue, to simplify reconstruction of the conductivity distribution to a 2D plane perpendicular to the longitudinal current.

The use of opposing electrode positions for drive current, where the two drive electrodes are on opposite sides of the sample, has been employed to reduce a current channelling effect present in the cerebrospinal fluid when imaging neural activity in the brain (Bayford 2006). We did not consider In-line electrode positions for drive current, where the two drive electrodes are on the same circumferential position and different electrode rings, as we predict a similar effect in the saline fluid layer of our model. We considered one possible longitudinal drive current pattern as in Fig 4: Opposing drive pattern, where drive current flows between two electrodes on opposite sides of the nerve and different electrode rings, generating a current with longitudinal components between the two electrode rings. In contrast to the convention in (Graham and Adler 2007), to record boundary voltages as differential measurements between electrode pairs configured in the same way as the drive current electrodes, we implemented our measurement electrode pairs in an in-line pattern in order to record the longitudinal boundary voltage gradients across the imaging volume. To abbreviate we refer to this electrode pattern with opposing drive electrode pair and in-line measurement pairs as 'Opposing-in-line'.

The drive and measurement electrode pattern to use for EIT modelling was selected to maximise the sensitivity to changes in conductivity in the 9 central grids (3 x 3 squares) which produce the lowest changes in boundary voltages and so, accordingly, are most difficult to detect (Holder 2004). Sensitivity was defined as the standard deviation of the normalised change in boundary voltages (Fan, Wang et al. 2015), in our case in response to conductivity changes across the set of 9 central grids.

The forward problem was solved using the single fascicle FEM model with each of the 49 sub volumes individually set to an active state. A 10 µA amplitude current, in accordance with (Fouchard, Coizet et al. 2017), was applied across the drive electrodes. In the 32 electrode nerve cuff there are a total of 16 possible drive electrode pairs for each longitudinal current pattern. We applied a DC drive current and assumed negligible impedance at the electrode-tissue interface.

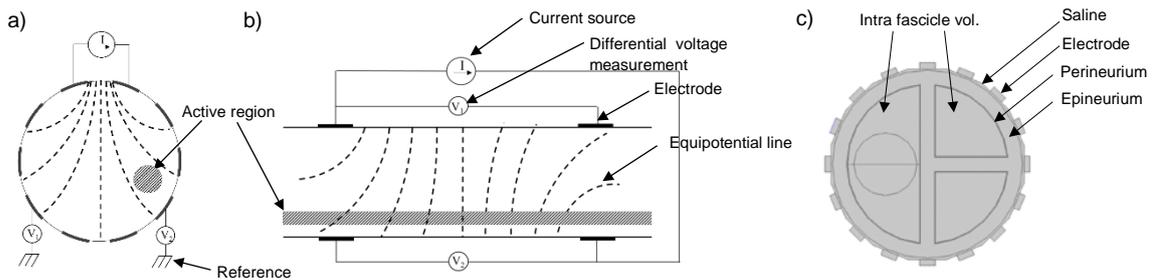

Figure 4: Section views of a possible transverse current pattern (a), and Opposing-in-line electrode pattern (b), with one opposing drive electrode pair and two in-line measurement electrode pairs visible. End view of the three fascicle model used to generate pseudo-experimental data (c).

For the inverse EIT problem we used Zeroth order Tikhonov regularisation to invert the sensitivity matrix as was done in neural-EIT experiments in (Aristovich, Santos et al. 2014, Vongerichten, dos Santos et al. 2016). The Tikhonov solution to the general inverse equation $\mathbf{c}_\tau = \mathbf{A}^\dagger \mathbf{g}$ is given by:
$$\mathbf{c}_\tau = \mathbf{VF\Sigma}^\dagger \mathbf{U}^T \mathbf{g} \tag{5}$$
where $\mathbf{V}$ and $\mathbf{U}$ are orthogonal matrices from singular value decomposition (SVD) regularisation of the sensitivity matrix $\mathbf{A} = \mathbf{U\Sigma V}^*$, $\mathbf{\Sigma}$ is the diagonal matrix of singular values $s_i$, $*$ indicates the Hermitian, and $\mathbf{F}$ is a diagonal matrix where $\mathbf{F}_{i,i} = s_i^2/(s_i^2 + \tau^2)$. The L-curve, or Pareto frontier curve, method was used as an unbiased means to select the Tikhonov regularisation parameter $\tau$ (Nasehi Tehrani, McEwan et al. 2012, Aster, Borchers et al. 2013).

Unbiased parameter selection is one of several recommendations made in (Holder 2004) to avoid committing an 'inverse crime' when using simulated instead of real data in inverse problems. Other recommendations that were implemented are: 1) Mesh size: our forward problem was meshed with a max mesh size of 4mm, a min mesh size of 0.1mm, a max growth rate of 2, a curvature factor of 0.4, and a resolution of narrow regions of 0.7. Our simulated data were acquired with the max and min mesh sizes reduced to 2mm and 0.05mm respectively. 2) Shape of conductivity anomalies: we generated simulated data on a three-fascicle model of nerve containing one semi-circular fascicle and two smaller quarter-circle shaped fascicles of equal size, Fig 4c. Each fascicle was encompassed by a 5 µm thick perineurium layer and separated by a 90 µm thick epineurium layer. Thus, the geometry and location of the epineurium, perineurium and intra-fascicle tissues in the three-fascicle model differs from the single fascicle model. However, in both models a 70 µm thick saline fluid layer, of 2 S/m conductivity, separated the electrodes and nerve outer boundary, and the epineurium-saline boundary had the same size and location. 3) Simulating noise: white noise of +/-35µV$_{RMS}$ amplitude and Gaussian distribution was added to each simulated voltage measurement. 4) Hardware errors: a number within the range +/- 100 µV was randomly selected and added to each electrode pair as a fixed offset, which was intended to account for accuracy of the ADC hardware and electrode impedance measurement errors. 5) Quantization: all values were rounded to the nearest multiple of 10µV.

## III. RESULTS

Table 2: Electrical properties values used in unmyelinated and myelinated nerve fibre models.

| Parameter | Fixed value | Maximum | Minimum | Units |
|---|---|---|---|---|
| **Unmyelinated** | | | | |
| Membrane conductivity | | 0.0008 | 0.000144 | S/cm$^2$ |
| Membrane specific capacitance | 0.81 | | | uF/cm$^2$ |
| **Myelinated** | | | | |
| Membrane conductivity in the regions | | | | |
| Node of Ranvier | | 1.95 | 0.018 | S/cm$^2$ |
| Paranode | 0.001 | | | S/cm$^2$ |
| Juxtaparanode | | 0.0082 | 0.002 | S/cm$^2$ |
| Internode | | 0.00269 | 0.002 | S/cm$^2$ |
| Membrane specific capacitance | 2 | | | uF/cm$^2$ |
| Myelin specific capacitance per lamella (2 membranes) | 0.05 | | | uF/cm$^2$ |
| Myelin conductivity per lamella (2 membranes) | 0.0005 | | | S/cm$^2$ |
| **Fluid and extracellular volumes** | | | | |
| Intracellular fluid resistivity | 70 | | | Ω.cm |
| Periaxonal fluid resistivity | 70 | | | Ω.cm |
| Extracellular space resistivity | 1,000 | | | Ω.cm |
| Extracellular space volume fraction (with myelin volume excluded) | 0.5 (0.677) | | | |

*A. Membrane Conductivity*

Membranes dynamics modelling showed that during an action potential both the relative increase in membrane conductivity and the maximum value reached are significantly higher in myelinated fibres, particularly at the node of Ranvier, than in unmyelinated fibres, see values in table 1. Higher membrane conductivity is attributable to higher concentration of ion channels and facilitates a higher transmembrane ion flux, which, evidently, for Sodium ions was estimated to peak at around -33 mA/cm$^2$ in the node of Ranvier compared to -0.045 mA/cm$^2$ in an unmyelinated fibre. In an unmyelinated fibre, the maximum membrane conductivity increases by a factor of 5.56 up to a maximum value of 0.0008 S/cm$^2$. In a myelinated fibre, the node of Ranvier produces both the highest increase in conductivity, a factor of 108, as well as the highest maximum conductivity value of 1.95 S/cm$^2$. In comparison, the juxtaparanode changes by a factor of 4.1 to reach a maximum value of 0.0082 S/cm$^2$, and the internode changes by a factor of 1.35 to reach a maximum value of 0.00269 S/cm$^2$. The membrane surface area of the node of Ranvier is 2 to 3 and 3 to 4 orders of magnitude smaller than the juxtaparanode and internode regions respectively. Taken in this context, the significantly higher membrane conductivity of the node of Ranvier is explained by necessity to enable a large current through a relatively small area, i.e. a high transmembrane ion flux, during an action potential.

*B. Transverse Fibre Impedance*

For a myelinated fibre of 10 µm diameter (and 3 to 15 µm range), fraction change in transverse impedance of -1.6x10$^{-10}$ (-4.6x10$^{-11}$ to -2.3x10$^{-10}$) was predicted in the juxtaparanode region, and -5.9x10$^{-12}$ (-1.8x10$^{-12}$ to -8.9x10$^{-12}$) in the internode region, Fig 5a & 5b. In contrast, in the node of Ranvier a fraction change of -1.4x10$^{-3}$ (-4.3x10$^{-4}$ to -2.0x10$^{-3}$) was predicted, which dominated the overall response of the fibre despite accounting for only 0.087% (0.33 to 0.069 %) of the total fibre volume, Fig 5c & 5d. In the myelinated regions, the high resistance myelin sheath acts to shield the axon from the extracellular current. At frequencies above approximately 10$^9$ Hz at which capacitive current traverses the myelin sheath, capacitive current also traverses the axon membrane, producing negligible changes in resistivity at all frequencies and rendering changes in the axon membrane conductivity essentially undetectable. The unmyelinated node of Ranvier, on the other hand, shows a reasonable change in impedance magnitude below approximately 10$^6$ Hz, above which capacitive current traverses the axon membrane. An error of unknown magnitude will arise from treating the fibre regions as separate 2D entities that are only connected by longitudinal current pathways in the extracellular space, as opposed to a 3D model with several transverse and longitudinal current pathways.

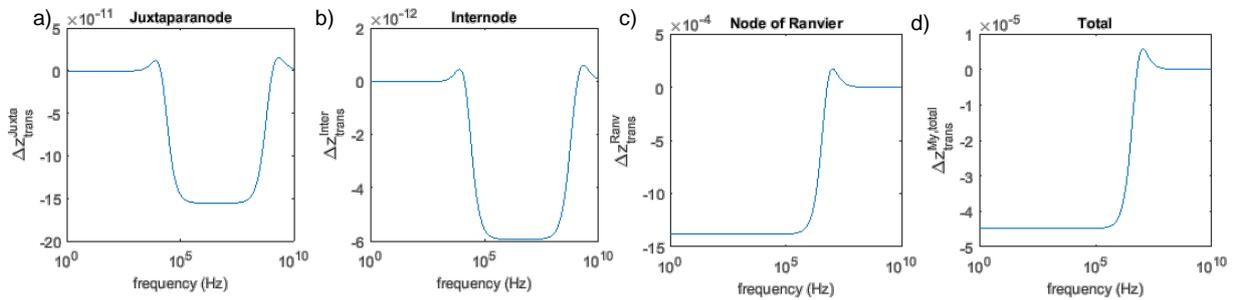

Figure 5: Frequency response of the fraction change in transverse impedance for the juxtaparanode (a), internode (b), and node of Ranvier (c) regions for a 15 µm diameter myelinated fibre, and their combined transverse impedance corresponding to half of a node-node length (d). The node of Ranvier dominates the overall response despite accounting for the smallest volume fraction of the fibre.

The fraction change in transverse impedance of an unmyelinated fibre of 1.5 µm diameter was estimated to be -1.4x10$^{-4}$ at frequencies below 10$^4$ Hz. A frequency drop off, due to capacitive current traversing the axon membrane, begins at 10$^5$ Hz, an order of magnitude lower than for myelinated fibres, and settles at approximately 5x10$^7$ Hz after an incursion into positive fraction changes in impedance which peaks at 5.8x10$^{-5}$ at 1.7x10$^6$ Hz. The fraction change in transverse impedance in unmyelinated fibres is larger, by a factor of 3 or more, than all modelled diameters of myelinated fibres with the exception a small frequency range (5x10$^5$ to 10$^6$ Hz) where the curves intersect, Fig 6. Outside of this narrow frequency range activity in unmyelinated fibres would disproportionately affect detected impedance changes in intra-fascicle tissue when using transverse current in neural-EIT.

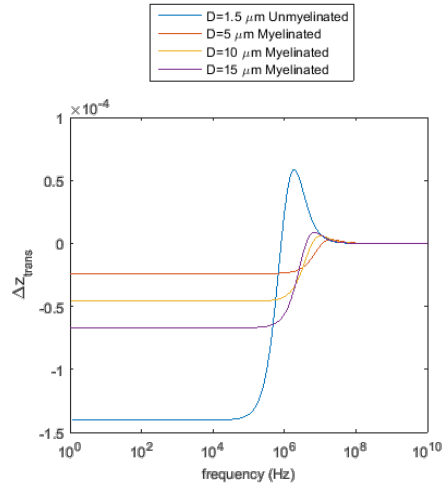

Figure 6: Frequency response of the fraction change in transverse impedance for unmyelinated and myelinated fibres. The fraction change is larger in unmyelinated fibres than in myelinated fibres at all frequencies except for the range $5\times10^5$ to $10^6$ Hz

*C. Longitudinal Fibre Impedance*

For longitudinal fibre orientation, *LTSpice* software was used to compare the lumped parameter model and the physiological model using absolute resistance and capacitance component values for a single length of myelinated fibre extending between two neighbouring nodes of Ranvier. At frequencies below $10^4$ Hz the lumped parameter model produced an impedance magnitude which was higher than that of the physiological model by a factor of 1.4 to 1.9 in the inactive state and by a factor of 1.01 to 1.03 in the active state, Fig 7. There is a noticeable discrepancy between the two models in the shape of the curves at around $10^4$ Hz, where the physiological model contains two additional inflection points closely spaced together, which can be attributed to capacitive current crossing the axon membrane in the myelinated regions. The observed differences between the two models are caused by current entering the periaxonal space longitudinally between the node of Ranvier and paranode regions, particularly in the inactive state. We believe that the value used for the resistance which controls longitudinal current between the node of Ranvier and paranode regions is too high. This belief is supported qualitatively by accumulation of potassium ions and swelling in the paranodes during sustained activity observed in (Geada Trigo Calheiros De Figueiredo 2014).

Equation 4 predicted significant differences in the frequency response between unmyelinated fibres and the lumped parameter model of myelinated fibres as seen in Figs. 8a and 8b. In unmyelinated fibres and 10mm electrode spacing, the largest fraction change was -0.10 at 1 Hz. The frequency drop off caused by capacitive current across the axon membrane began at 1Hz and settled around at 300 Hz after an incursion into the positive range, which peaked with a value of $3.9\times10^{-3}$ at 100 Hz. In myelinated fibres, for all diameters the frequency drop off began at around 200Hz and settled to zero at 1MHz after an incursion into positive values beginning at around 100 kHz. Higher frequency drop off values in myelinated fibres are expected due to the presence of the lower capacitance myelin sheath. The largest fraction change for all myelinated fibre diameters were observed below 200 Hz. The magnitude of the fraction change increased with decreasing fibre diameter, and with decreasing electrode spacing down to 2 mm. The frequency at which the frequency drop off begins changed negligibly across different fibre diameters and electrode spacing. The difference in the frequency drop off between unmyelinated and myelinated fibres indicates it is possible to screen out unmyelinated fibre activity using the operating frequency, and to distinguish their activity using multi-frequency EIT.

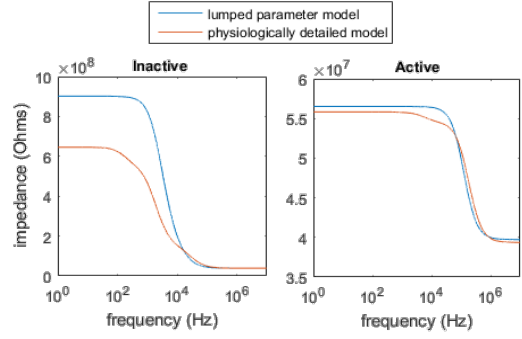

Figure 7: Impedance of the lumped parameter model and physiologically detailed model for a myelinated fibre in inactive (left) and active (right) states. There is a significant difference between the two models in the inactive fibre state at frequencies below $10^4$ Hz, and good agreement between the two models in the active state at all frequencies.

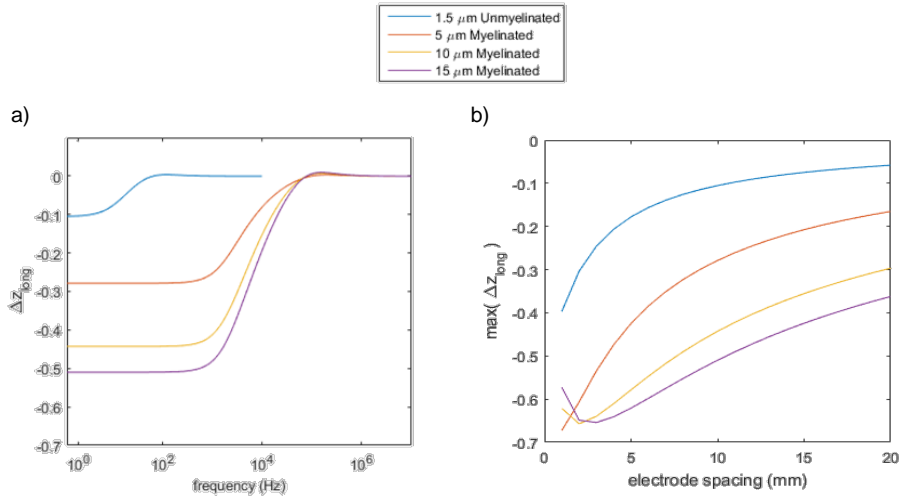

Figure 8: Frequency response of the fraction change in longitudinal impedance for unmyelinated and lumped parameter model of myelinated fibres at 10 mm electrode spacing (a). The frequency drop off occurs at much lower frequency in unmyelinated fibres than in myelinated fibres. The maximum fraction change in longitudinal impedance, which occurs at 1 Hz, shows dependence on electrode spacing and fibre diameter (b).

*D. Nerve Tissues*

The time constants of a 1.5 µm diameter unmyelinated fibre were 2.5 ms in the active state and 14 ms in the inactive state; producing a capacitance correction factor $T_p$ of 2.8. In contrast, lumped parameter model values of a 10 µm diameter myelinated fibre produced time constants of $1.61 \times 10^{-2}$ ms in the active state and 1.05 ms in the inactive state, with little variation from these values for other fibre diameters; $T_p$ was 1.00. In the velocity correction factor $C_p$, transit time effects through the nerve cuff were found to be negligible due to the high signal velocities in myelinated fibres and significant drop in resistance at the active nodes of Ranvier. Signal dispersion effects reduced the fraction change in impedance by a factor of 1.16 over 30 mm travel distance, increasing to a factor of 1.94 over 100mm. Dispersion correction is most relevant to an experimental set-up with controlled nerve excitation, where the travel distance is known.

At 1 kHz the estimated impedance magnitude of intra-fascicle tissue reduced from 1,142 Ω.cm (inactive) to 1,141 Ω.cm (active) in the transverse orientation, a fraction change of $-8.8 \times 10^{-4}$, and from 328 Ω.cm (inactive) to 230 Ω.cm (active) in longitudinal orientation, a fraction change of -0.30. For comparison, impedance values reported in (Ranck Jr and BeMent 1965) from in-vivo measurements on the dorsal column of cat, at 1 kHz and 10 mm electrode spacing, are 1,211 Ω.cm transverse and 138 to 217 Ω.cm longitudinal. The peripheral nerve is not directly comparable to the dorsal column because the latter is part of the central nervous system and so contains neurones and additional types of glial cells (Pannese 2015). In addition, the experiments on dorsal column were performed in-vivo, under sodium barbiturate anaesthetic, and so may include both active and inactive neurones and nerve fibres.

*E. Electrical Impedance Tomography*

Sensitivity analysis of the electrode patterns produced standard deviations across the normalised boundary voltage measurements of $2.63 \times 10^{-5}$ for transverse current pattern with drive current and voltage measurement between adjacent electrodes, and $5.89 \times 10^{-5}$ for Opposing-in-line pattern at 10mm electrode spacing increasing to $1.66 \times 10^{-4}$ at 2mm electrode spacing; we therefore selected Opposing-in-line pattern.

In Opposing-in-line pattern, due to the increased resistance between the two current electrodes, both the voltage across the drive current electrodes and the signal to error ratio (SER), which is the combined effect of the upper limit of the assumed noise and errors and is calculated using normalised values, increases with increasing electrode spacing. The SER of the normalised change in differential voltages increased by a factor of 10 between electrode spacings of 2mm and 10mm, and by a factor of 2 between electrode spacings of 6mm and 10mm. Referring back to Fig 8, when selecting the electrode spacing a trade-off exists between the fraction change in impedance and the SER. Incidentally, in neural EIT experiments researchers have typically averaged recordings across multiple measurements to reduce noise (Aristovich, Dos Santos et al. 2015, Aristovich 2016, Vongerichten, dos Santos et al. 2016, Fouchard, Coizet et al. 2017). Averaging cannot be implemented in real-time imaging, a crucial requirement for a neural prosthetics control interface, and so has not been assumed in our model.

With 10mm electrode spacing EIT of activity in each fascicle of the three fascicle model were clearly distinguishable from one another by the levels of reconstructed activity in different sub-volumes, Fig 10. SERs in the normalised change in differential voltages were 24 in semicircle shaped fascicle 1, and 18 in the two quarter-circle shaped fascicles 2 and 3, indicating real-time imaging of fascicle level activity is achievable in experiments on rat sciatic nerve at the assumed noise and error levels. EIT of sub-fascicle activity within fascicle 1 reconstructed different magnitudes of activity in the same sub-volumes. The limited intra-fascicle resolution may be improved by a higher number of sub-volumes used in the forward solution, reduced noise and errors, inclusion of passive tissue structures in the forward solution (Jehl, Aristovich et al. 2016), and use of additional reconstruction steps in solving the inverse problem (Aristovich, Santos et al. 2014).

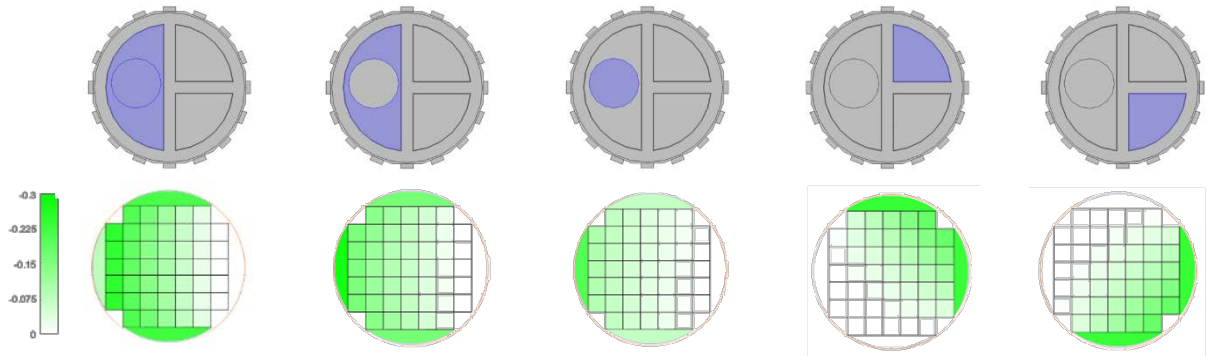

Figure 10: Activity in the three fascicle model (top row), with purple indicating -0.286 fraction change in impedance, and corresponding reconstruction using the EIT algorithm, which was populated with data from the single-fascicle model, (bottom row) with colour scale showing fraction change in impedance.

IV. DISCUSSION

The inclusion of passive tissues and a saline fluid layer in our model, to better describe the response of a whole nerve, reduces the fraction change in impedance. The extent of this reduction depends on the thickness of the passive tissue and saline fluid layers. For comparison, results of impedance experiments on rat sciatic nerve by Ref (Fouchard, Coizet et al. 2017) reported an undetectable fraction change in transverse impedance, and a frequency dependent longitudinal fraction change in impedance with maximum value of -0.2 and frequency drop off beginning at approximately 200 Hz, see Fig 9. These experiment results are in reasonable agreement with our model prediction. More experiment data is required for a direct comparison, some of which, e.g. saline fluid layer thickness, cannot be easily determined.

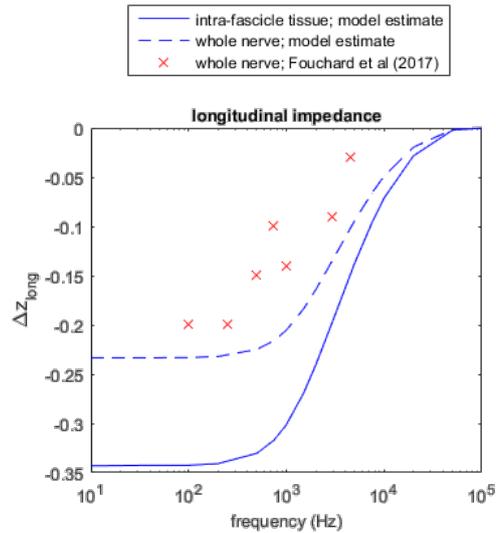

Figure 9: The fraction change in longitudinal impedance is lower for the whole nerve than for the intra-fascicle tissue as the whole nerve includes passive tissues and saline layer. Shape and magnitude of the curve for whole nerve shows reasonable agreement with Fouchard et al (2017).

For control of neural prosthetics, we may consider a scenario where only motor command fibres are active in each sub-volume of intra-fascicle tissue. Applying activity only in motor-command fibres to equation 5, using the statistical classification of fibres in (Prodanov 2006), reduces the expected fraction change in longitudinal impedance of intra-fascicle tissue to a maximum of -0.14 at 1 Hz (-0.12 at 1 kHz). Assuming an average asynchronous firing rate of 25 Hz across all motor fibres in an intra-fascicle tissue sub-volume, taken from motor unit firing rates associated with movement in the upper limb (Farina, Rehbaum et al. 2014), the expected fraction change in longitudinal impedance is -0.007 at 1 Hz (-0.006 at 1 kHz). While these magnitudes are relatively small they are not prohibitive to EIT imaging, in fact, they are comparable to fraction impedance changes observed in neural tissue in the cerebral cortex (unmyelinated grey matter) of rat of between -0.01 (DC) and -0.001 (10kHz) (Oh, Gilad et al. 2011) which have been successfully reconstructed using neural EIT by (Aristovich, Packham et al. 2016). Somatotopic organisation of fibres within fascicles would increase the magnitude of impedance change.

We have shown that EIT within a dual-ring electrode array operating with an Opposing-in-line pattern can detect fascicle level activity in a 3 fascicle model of rat sciatic nerve with SERs sufficient for real-time imaging. For neural prosthetics, reconstruction of fascicle level activity with an estimated fraction change in impedance of -0.007, representing activity in motor command fibres with a firing rate of 25 Hz, required a reduction in the noise and error values by a factor of 15 to 20. Reconstruction of fascicle level activity in all fascicles of the human median nerve, containing e.g. 10 fascicles at the elbow, would be a significant improvement in performance of nerve cuffs.

## V. CONCLUSION

In this study we considered the application of EIT to a nerve cuff for control of a robotic prosthetic by developing a model and then testing it using pseudo-experimental data synthesised to replicate real operating conditions. The results in this study provide an estimate of the change in impedance of intra-fascicle tissue in mammalian nerve, and present a viable EIT drive and measurement electrode pattern, implemented on a dual-ring electrode array, to detect impedance changes in the anisotropic axis. These results are necessary steps towards implementing EIT in a nerve cuff and show extreme promise for developing advanced neural prosthetics interfaces.